\documentclass[reprint,longbibliography,
superscriptaddress,
amsmath,amssymb,
aps,
prl,
floatfix,
]{revtex4-2}

\usepackage{graphicx}
\usepackage{amsmath}
\usepackage{amsfonts}
\usepackage{color}
\usepackage[normalem]{ulem}
\usepackage[colorlinks, linkcolor= blue, citecolor = blue, urlcolor=blue]{hyperref}

\begin{document}

\title{Harnessing the natural resonances of  time-varying dispersive interfaces}
\author{Carlo Rizza} 
\affiliation{Department of Physical and Chemical Sciences, University of L'Aquila, I-67100 L'Aquila, Italy}
\author{Maria Antonietta Vincenti}
\affiliation{Department of Information Engineering, University of Brescia, I-25123 Brescia, Italy}
\author{Giuseppe Castaldi}
\affiliation{Department of Engineering, Fields \& Waves Lab, University of Sannio, Benevento, I-82100, Italy} 
\author{Alessandra Contestabile}
\affiliation{Department of Physical and Chemical Sciences, University of L'Aquila, I-67100 L'Aquila, Italy}
\author{Vincenzo Galdi} \email{vgaldi@unisannio.it}
\affiliation{Department of Engineering, Fields \& Waves Lab, University of Sannio, Benevento, I-82100, Italy}
\author{Michael Scalora}
\affiliation{Aviation and Missile Center, U.S. Army CCDC, Redstone Arsenal, Alabama 35898-5000, USA}

\begin{abstract}
Space–time modulation of electromagnetic parameters offers novel exciting possibilities for advanced field manipulations. In this study, we explore wave scattering from a time-varying interface characterized by a Lorentz-type dispersion with a step-like temporal variation in its parameters.  Our findings reveal a new process: an unconventional frequency generation at the natural resonances of the system. Remarkably, this phenomenon enables the coupling of propagating waves to evanescent ones, allowing the direct far-field excitation of surface-wave modes without the mediation of spatial gratings or prisms. These results suggest a novel strategy for designing compact and ultra-fast photonic devices, eliminating the necessity for subwavelength spatial structuring or prolonged temporal modulations.   
\end{abstract}

\maketitle
The growing accessibility of advanced metamaterial platforms has revitalized research efforts in the study of time-varying media. The ability to manipulate their constitutive properties both in space and time (see, e.g. \cite{Cui:2024ro}, for a recent roadmap) has been a topic of longstanding interest in electrodynamics, with seminal contributions dating back to the 1950s \cite{Morgenthaler:1958vm,Oliner:1961wp,Felsen:1970wp,Fante:1971to}.

Recent studies have investigated various approaches aimed at leveraging the {\em temporal} structuring of metamaterials, either independently or in conjunction with spatial configurations, to enable advanced field manipulations \cite{Caloz:2020sm1,Caloz:2020sm2,Engheta:2021mw,Pacheco:2022tv}, and overcome typical limitations of linear, time-invariant systems \cite{Hayran:2023ut}. Examples of such explorations include analogies like temporal boundaries \cite{Xiao:2014ra} and interfaces \cite{Stefanini:2022ti}, total internal reflection \cite{Bar-Hillel:2024tr}, photonic time crystals \cite{Romero:2016tp,Lyubarov:2022ae}, time gratings \cite{Galiffi:2020}, and temporal impedance matching \cite{Pacheco:2020at,Castaldi:2021es,Galiffi:2022tp}, among others. 
Although challenging to realize, several studies have proposed various platforms for achieving time-varying media with ultrafast modulation of the electromagnetic response \cite{Lee:2018,Zhou:2020,Bohn:21,Moussa:2023,Lustig:2023,Tirole2023}.  For a comprehensive overview of this rapidly evolving research field, readers are directed to \cite{Galiffi:2022po} and references therein.

In this context, time-varying metasurfaces have emerged as a technologically viable platform to harness spatio-temporal modulations, enabling a myriad of effects and phenomena such as frequency conversion and generation \cite{Shcherbakov:2019pa,Karl:2020fc,Taravati:2021pa,Liu:2021pa}, time-varying optical vortices \cite{Sedeh:2020}, polarization converters \cite{Hu:2022}, unconventional wave diffraction \cite{Amor:2023}, and
broadband spectral camouflage \cite{Liu:2019}. A remarkable advantage offered by these platforms
 is their ability to generate new frequencies within compact designs, bypassing the need for conventional nonlinear effects \cite{Lee:2018}. 
Of particular relevance to our current investigation is the concept of a {\em time-grating} in a time-varying thin film \cite{Galiffi:2020}. This entails a time-periodic modulation of an impedance surface that sustains guided modes, and provides an alternative approach to coupling propagating waves with evanescent ones. Unlike conventional gratings, it does not require subwavelength spatial structuring.

\begin{figure}
	\centering
	\includegraphics[width=\linewidth]{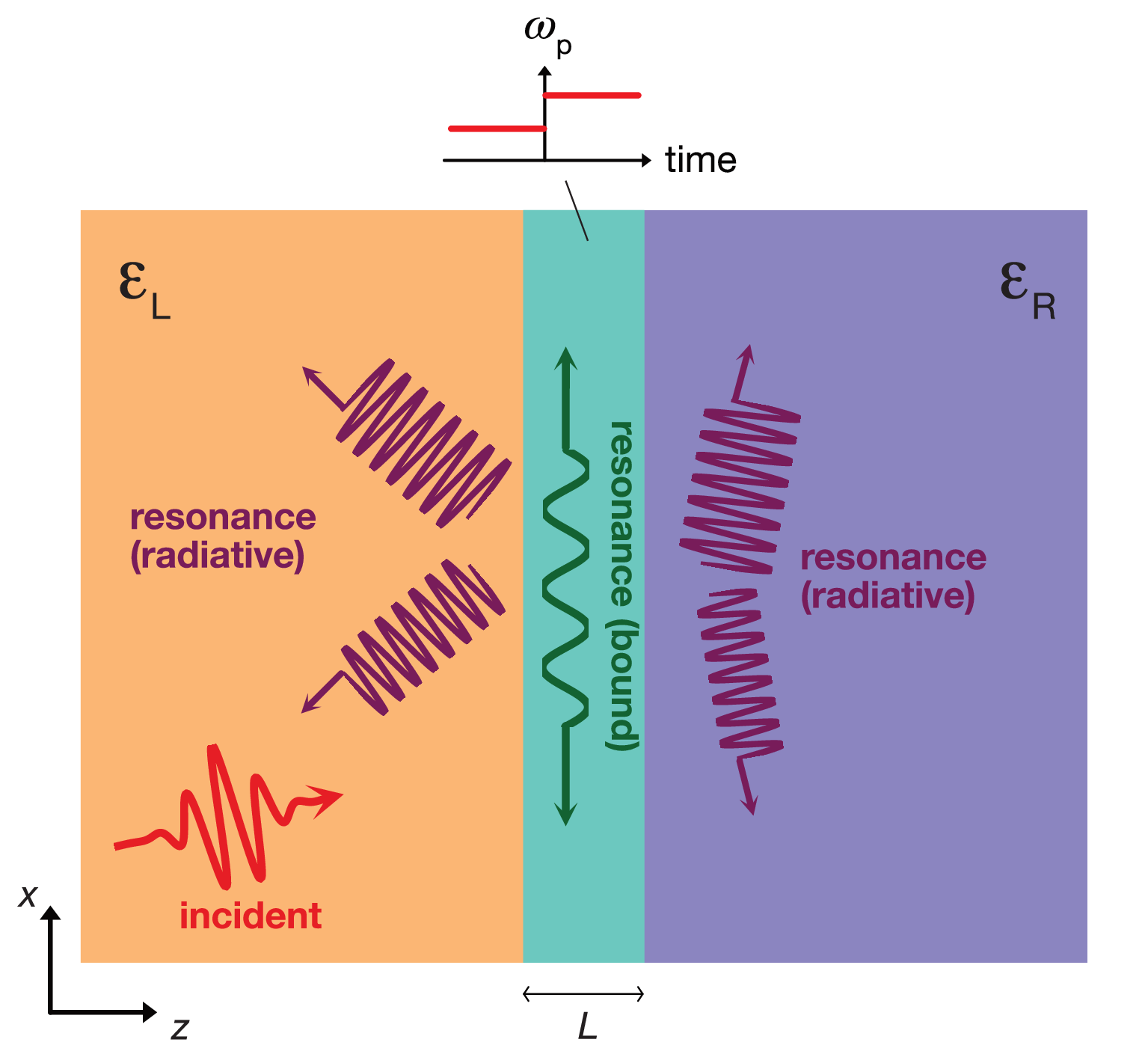}
	\caption{Problem schematic. A time-varying dispersive interface, in the form of a thin dielectic layer, is sandwiched between two homogeneous, stationary half-spaces with  relative permittivities $\varepsilon_L$ and $\varepsilon_R$. The interface undergoes a temporal boundary (i.e., a sudden switch of the plasma frequency $\omega_p$) which, upon interaction with an incident wave packet, can excite either radiative or bound resonant modes.}
	\label{Figure1}
\end{figure}
In this study, we explore a time-varying, dispersive  interface that features Lorentzian dispersion, whose parameters undergo temporal modulation. We find that this structure can exhibit an unusual linear frequency generation at the resonance, which in turn enables coupling between propagating and evanescent waves, without the need for either periodic or extended temporal modulations.   Instead, these phenomena are driven by a temporal boundary effect, i.e., a sudden step-like change in the interface's constitutive properties. We demonstrate that such a temporal boundary can trigger the inherent natural resonances of the time-varying dispersive interface. 
As a result, depending on whether these natural resonances correspond to radiative or bound modes, the proposed configuration can act either as a frequency converter or enable the far-field excitation of surface waves.

As schematically illustrated in Fig. \ref{Figure1},  we consider a time-varying, dispersive dielectric layer of thickness $L$ sandwiched between two homogeneous, stationary and nondispersive half-spaces, with relative permittivity $\varepsilon_L$ and $\varepsilon_R$. Although our proposed mechanism may be observed in layers of arbitrary thickness, to simplify our analytical approach we assume that the central layer is very thin, i.e., $L\ll\lambda$, with $\lambda$ denoting a characteristic wavelength \cite{Stumpf:2023}, and ideally treat it as a zero-thickness metasurface (located at $z=0$). These conditions will be relaxed later on. Our analysis is focused on transverse electric (TE) wave packets with electric field  given by $\vec{\cal{E}}(x,z,t)=\textrm{Re} \left[E(z,t)e^{i k_x x} \right] \hat{\bf e}_y$. Here and henceforth, $k_x$ denotes the conserved transverse wavenumber, $\hat{\bf e}_{\alpha}$ is an ${\alpha}$-directed unit vector, and $i$ stands for the imaginary unit. Moreover, we define a generic field quantity in the form ${\cal{G}}(x,z,t)=\textrm{Re} \left[G(z,t)e^{i k_x x} \right]$. For such scenarios, the interaction with the metasurface may be effectively described in terms of sheet transition conditions,
\begin{equation}
\label{ST}
 \Delta E(z,t) = 0, \quad \Delta  H(z,t) =   \frac{d P(t)}{d t}, 
\end{equation}
with $\Delta G (z,t)=G(z=0^+,t)-G(z=0^-,t)$ ($G=E,H$) denoting the discontinuity across the metasurface of the reduced tangential electric and magnetic field components \cite{Achouri:2015}, and $P(t)$ the $y$-component of the surface polarization. The metasurface is modeled to exhibit a time-varying Lorentzian dispersion, characterized by a temporal variation in its plasma frequency. As a result, the temporal evolution of the surface polarization $P(t)$ is governed by the following second-order differential equation
\begin{equation}
\label{Lor_1}
   \frac{d^2 P(t) }{dt^2}+\gamma \frac{d P (t) }{d t}+\omega_0^2 P(t)=\varepsilon_0  L \omega_p^2(t) E (0,t), 
\end{equation}
where $\varepsilon_0$ is the vacuum permittivity,
$\omega_0$ and $\gamma$ and the resonance frequency and damping parameter, respectively (which remain constant over time),
and $\omega_p(t)$ is the time-varying plasma frequency.
Equation (\ref{Lor_1}) models a broad category of time-varying media with an effective Lorentzian dispersion (either thin homogeneous layers of natural materials or, more generally, metasurfaces) where the temporal modulation of $\omega_p$ may be induced by changing  the volumetric density of polarizable atoms or meta-atoms \cite{Solis:2021,Horsley:2023}. For instance, the plasma frequency of a semiconductor-based metasurface can be dynamically changed in real-time through processes such as photocarrier injection \cite{Lee:2018}.

In the stationary scenario (i.e., constant plasma frequency), the metasurface exhibits a continuous spectrum of radiative modes that can naturally couple with propagating waves in the surrounding halfspace, along with the potential for discrete bound modes (surface waves) that propagate along $x$ and exhibit evanescent decay along $z$. These bound modes cannot be directly excited by a propagating wave impinging from either side, and either prism- or grating-type schemes are typically employed to bridge the wave vector gap. As demonstrated in Ref. \cite{Galiffi:2020}, a {\em periodic} time modulation of the susceptibility offers an alternative approach. Here, we take a different route, assuming an {\em abrupt} change over time of the plasma frequency in Eq. (\ref{Lor_1}). Specifically, setting $\omega_p^2(t)=A(t) \omega_0^2$, we consider
\begin{equation}
\label{AA}
A(t)=A_1  +(A_2-A_1)  U (t),
\end{equation}
with $U\left(t\right)$ denoting the conventional unit step ($=1$ for $t>0$, and $0$ otherwise).  As it will be demonstrated in the following discussion, this temporal boundary can trigger the natural resonances supported by the metasurface when interacting with an incident wave packet. These resonances can manifest as either radiative or bound modes, which can couple with a propagating incident field that is significantly detuned from the resonance.

It is worth noting that previous studies of time-varying metasurfaces either disregarded dispersion entirely \cite{Stefanini:2022ti, Wang:2023cs} or did not consider abrupt changes  \cite{Tiukuvaara:2021,Amra:2024}. Notable exceptions are found in Refs. \cite{Lee:2018,Wang:2023mb,Duan:2024,Guizhen:2024}, where rapidly switched dispersive metasurfaces are investigated.
 Although these studies successfully demonstrate linear frequency conversion, their underlying assumptions hinder the excitation of the natural resonances associated with the specific dispersion profile.

By solving Maxwell's equations in conjunction with the sheet transition conditions (\ref{ST}), as well as Eqs. (\ref{Lor_1}) and (\ref{AA}), we derive analytical solutions for the TE wave-scattering scenario schematized in Fig. \ref{Figure1} (see \cite{supp} for more details).

To gain insights into the underlying physical mechanism, we focus on the temporal dynamics of the surface polarization $P(t)$, which may be expressed as
\begin{equation}
\label{PPP}
P(t)= U(-t)P_1(t)+ U(t) [P_m (t)+P_2 (t)].
\end{equation}
In Eq. (\ref{PPP}), $P_1(t)$ and $P_2(t)$ denote the steady-state surface polarizations corresponding to $A=A_1$ and $A=A_2$, respectively, and may be expressed in terms of the incident electric field as follows:
\begin{eqnarray}
\label{P_k}
P_{1,2}(t)= \varepsilon_0 \int_{-\infty}^{+\infty} d \omega e^{-i \omega t} {\tilde F}_{1,2} (\omega) {\tilde E}_i(\omega), 
\end{eqnarray}
where ${\tilde E}_i(\omega)$ is the Fourier spectrum of the incident field, and ${\tilde F}_{1,2} (\omega)$ are transfer functions whose analytical expressions are explicitly given in \cite{supp}.
The term $P_m(t)$, which is less immediately apparent, constitutes a crucial contribution stemming from the natural resonances associated with the dispersion profile and, more specifically, the steady-state TE eigenmodes supported by the interface at $t>0$ (i.e., $\omega_p^2=A_2 \omega_0^2$). For these modes, the reduced electric field can be expressed as $E(z,t) = E_m e^{-i(K_{z}^{(L)} z +\Omega t)} $, and  $E(z,t)=E_m e^{i (K_{z}^{(R)} z - \Omega t)}$ for $z<0$, and $z>0$, respectively, where $E_m$ is a constant amplitude, $K_{z}^{(j) 2}(k_x,\Omega)=\varepsilon_j \Omega^2/c^2-k_x^2$ ($j=L,R$), and $\Omega$ is a complex-valued angular eigenfrequency. In the low-loss regime (i.e., $\gamma \ll \omega_0$) and under the assumed thin-layer condition $L \ll \lambda$, the surface-polarization term associated with the above described eigenmodes can be approximated as 
\begin{eqnarray}
\label{P0}
P_m(t) \simeq P_+ e^{-i \Omega_+(k_x) t}+
P_- e^{-i \Omega_-(k_x) t}, 
\end{eqnarray}
where $P_{\pm}$ are complex-valued constants whose expressions are detailed in \cite{supp}, and 
\begin{equation}
\Omega_{\pm}(k_x) \simeq \pm \omega_0 \left[ 1-\frac{iL K_2(k_x,\pm \omega_0)}{2} \right]-i \frac{\gamma}{2},
\label{eq:DE}
\end{equation}
with $K_2(k_x,\pm \omega_0)=A_2 \omega_0^2/\{c^2 [K_{z}^{(L)}(k_x,\pm \omega_0)+K_{z}^{(R)}(k_x,\pm \omega_0)] \}$  (see \cite{supp} for further details). 
For a fixed $k_x$, defining the critical angular frequencies $\omega_L= k_x c/ \sqrt{\varepsilon_L}$ and $\omega_R= k_x c/ \sqrt{\varepsilon_R}$ in the media surrounding the time-varying interface, the resonant modes with angular eigefrequencies  $\Omega_{\pm} \simeq \pm \omega_0$ 
are {\em radiative}  [at angles $\theta \simeq \pm\arcsin {(\omega_j/\omega_0)}$ from the $z$-axis] if $\omega_0>\omega_{j}$, and {\em bound} otherwise. Accordingly, the term proportional to $K_2$ in Eq. (\ref{eq:DE}) is either purely  imaginary (consistently accounting for radiation losses) or real (implying a slight resonance shift), respectively. Note that negative frequencies are commonly observed in scenarios featuring a temporal boundary in a spatially unbounded medium \cite{Xiao:2014ra}, and physically correspond to backward waves since the wave vector is preserved.  In our assumed metasurface scenario, 
where only the $x$-component of the wave vector is preserved,
these negative frequencies are characterized by a backward (tilted) character along the $x$-direction.
In essence, the temporal boundary in Eq. (\ref{AA}) plays a key role in triggering the natural resonances of the time-varying interface. This occurs through the term $P_m(t)$, which arises from enforcing the continuity of the surface polarization and its first-order temporal derivative at $t=0$ \cite{Solis:2021}. This requirement is physically necessary, given Eq. (\ref{Lor_1}), even though the plasma frequency is discontinuous in time. The natural resonances  
can manifest as either radiative or bound modes, and can couple with a propagating wave packet impinging from the surrounding regions by suitably tuning the carrier frequency relative to the resonance one.

To elucidate the impact of the aforementioned phenomena on the electromagnetic dynamics, we explore two representative examples.
\begin{figure}
	\centering
	\includegraphics[width=\linewidth]{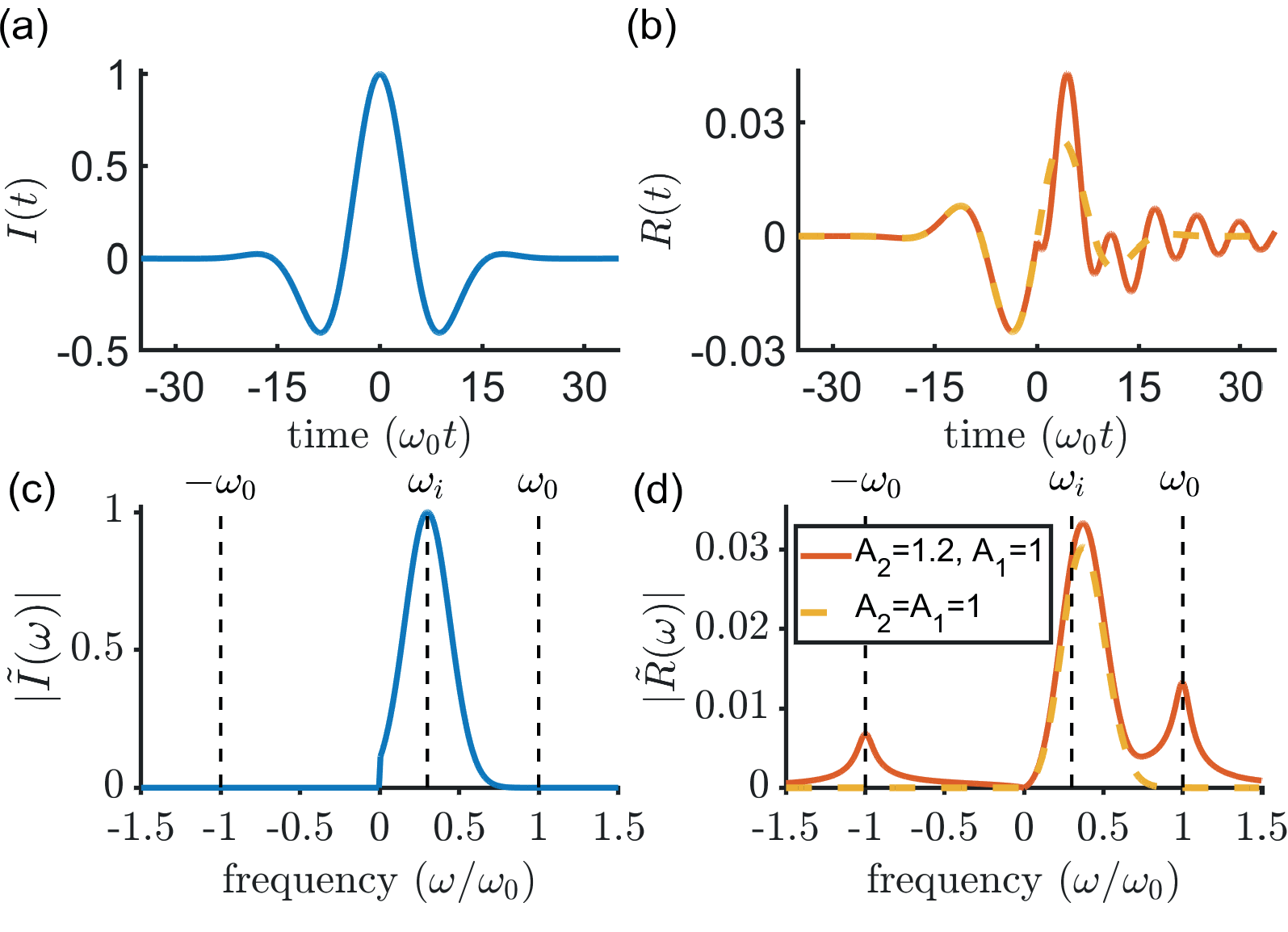}
	\caption{Example for normal incidence ($k_x=0$), with $\varepsilon_L=\varepsilon_R=1$, $A_1=1$, $L=0.16 c/\omega_0$, and $\gamma=5 \cdot 10^{-3} \omega_0$.
 (a), (b) Temporal profiles of the incident [$I(t)=E_i(0^-,t)/E_0$] and reflected [$R(t)=E_r(0^-,t)/E_0$] wave packets, respectively. 
 (c), (d) Corresponding spectra, with the dashed vertical lines indicating the positions of the carrier and resonance frequencies.
In panels (b) and (d), results in the presence ($A_2=1.2 A_1$) and absence ($A_2=A_1$) of the temporal boundary are compared. All spectra are normalized with respect to the peak value of incident one.}
	\label{fig2}
\end{figure}

In our initial example, we explore the potential for frequency generation induced by the time-varying interface. Referring to the scenario in Fig. \ref{Figure1}, in this case we assume a modulated Gaussian wave packet normally impinging (i.e., $k_x=0$) on the time-varying interface, with an electric field profile $E_i(0^-,t)=E_0 e^{-t^2/\sigma^2 - i\omega_i t}$. Here and henceforth, $E_0$ is a constant amplitude, $\omega_i$ is the carrier angular frequency, $\sigma$ controls the duration of the wave packet. The timing is selected so that the peak of the wave packet arrives at the metasurface precisely at the temporal boundary; the effects of imperfect synchronization will be addressed later. From the dispersion equation (\ref{eq:DE}), we observe that in this case the natural resonance manifests as a pair of degenerate radiative modes (at an angle $\theta=0$) with angular eigenfrequency $\Omega_{\pm}=\pm\omega_0$.
Figure \ref{fig2} illustrates the response of the system (in terms of reflected wave), assuming $\omega_i=0.3 \omega_0$, $\sigma=10/\omega_0$, $\varepsilon_L=\varepsilon_R=1$, and $L=0.16 c/\omega_0$, and $\gamma=5 \cdot 10^{-3} \omega_0$ with $c$ denoting the vacuum wavespeed. Specifically, Figs. \ref{fig2}a and \ref{fig2}c show the incident waveform and corresponding spectrum respectively, whereas Figs. \ref{fig2}b and \ref{fig2}d pertain to the reflected wave, considering the presence ($A_2=1.2A_1$; orange-solid curve) and absence ($A_2=A_1=1$; yellow-dashed curve) of the temporal boundary.
Remarkably, although the incident wave packet spectrum is negligible at the resonance angular frequency $\omega_0$, the reflected wave clearly exhibits significant components at $\pm\omega_0$ in the presence of a temporal boundary. As anticipated, this phenomenon is entirely absent in the stationary scenario (i.e., constant plasma frequency).
\begin{figure}
	\centering
	\includegraphics[width=\linewidth]{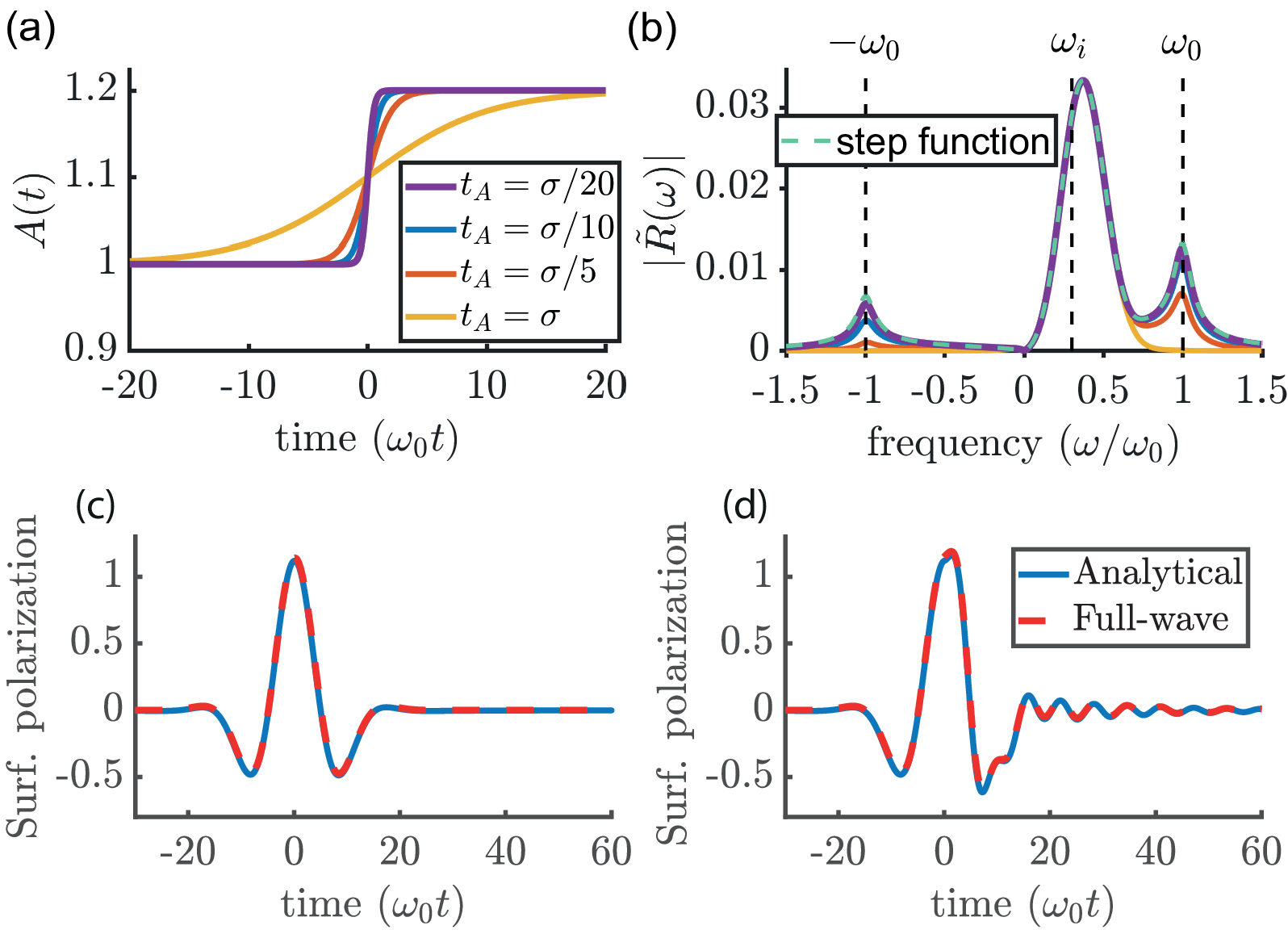}
	\caption{Configuration and parameters as in Fig. \ref{fig2}, but assuming a soft switching.
 (a), (b) Temporal profiles of the function $A(t)$ and corresponding spectra $|\tilde{R}(\omega)|$ of the reflected wave packet [$R(t)=E_r(0^-,t)/E_0$], respectively, for various values of the time scale $t_A$. In panel (b),  the response pertaining to a step variation is also shown (dashed curve) as a reference.
 (c), (d) Comparison between the analytically computed [${\cal{P}}_y(x=0,t)/(\varepsilon_0 E_0 L)$] and full-wave simulated [${\cal P}_y^{(s)}/(\varepsilon_0 E_0 L)$]  surface polarization 
 in the absence ($A_2=A_1$=1) and presence ($A_2=1.2 A_1$) of the temporal boundary, respectively. All spectra are normalized with respect to the peak value of incident one.}
	\label{fig3}
\end{figure}
For deeper understanding of the impact of the temporal boundary, we explore the scenario where the plasma frequency undergoes adiabatic switching with a characteristic time scale $t_A$. Similar scenarios have been explored in other time-varying configurations dealing with   temporal boundaries \cite{Galiffi:2022,Ganfornina-Andrades:23}.
For this generalized case, the problem is no longer solvable analytically, and we resort to numerical methods
(see \cite{supp} for details). In Fig. \ref{fig3}, we consider a plasma frequency variation given by $A(t)=0.5\left(A_2 +A_1\right)+0.5\left(A_2-A_1\right) \tanh\left(t/t_A\right)$, with various values of the time scale $t_A$ (see Fig. \ref{fig3}a). From the corresponding reflected-wave spectra shown in Fig. \ref{fig3}b, it is apparent that the solution obtained with $t_A=\sigma/20$ is in good agreement with our analytical predictions. As $t_A$ increases, the frequency-generation efficiency decreases, and the phenomenon virtually disappears for $t_A=\sigma$. 
For comprehensive validation,  we compare the analytical solutions [obtained assuming the step variation in Eq. (\ref{AA})] with full-wave simulations considering $t_A=\sigma/20$. The simulations involve a time-varying layer with thickness $L=0.16 c /\omega_0$, characterized by the volume-polarization $y$-component ${\cal P}_y^{(v)}(x,z,t)$, exhibiting Lorentz dispersion (see \cite{supp} for details). In Figs. \ref{fig3}c and \ref{fig3}d, we compare the temporal profiles of the analytical surface polarization $P(x=0,t)$ with the numerical one given by ${\cal P}_y^{(s)}= \int_{-L/2}^{L/2} d z {\cal P}_y^{(v)} (x=0,z,t)$, 
 in the presence and absence of the plasma frequency switch, respectively.
 Remarkably, the agreement is very good in both cases.

\begin{figure*}
	\centering
	\includegraphics[width=\linewidth]{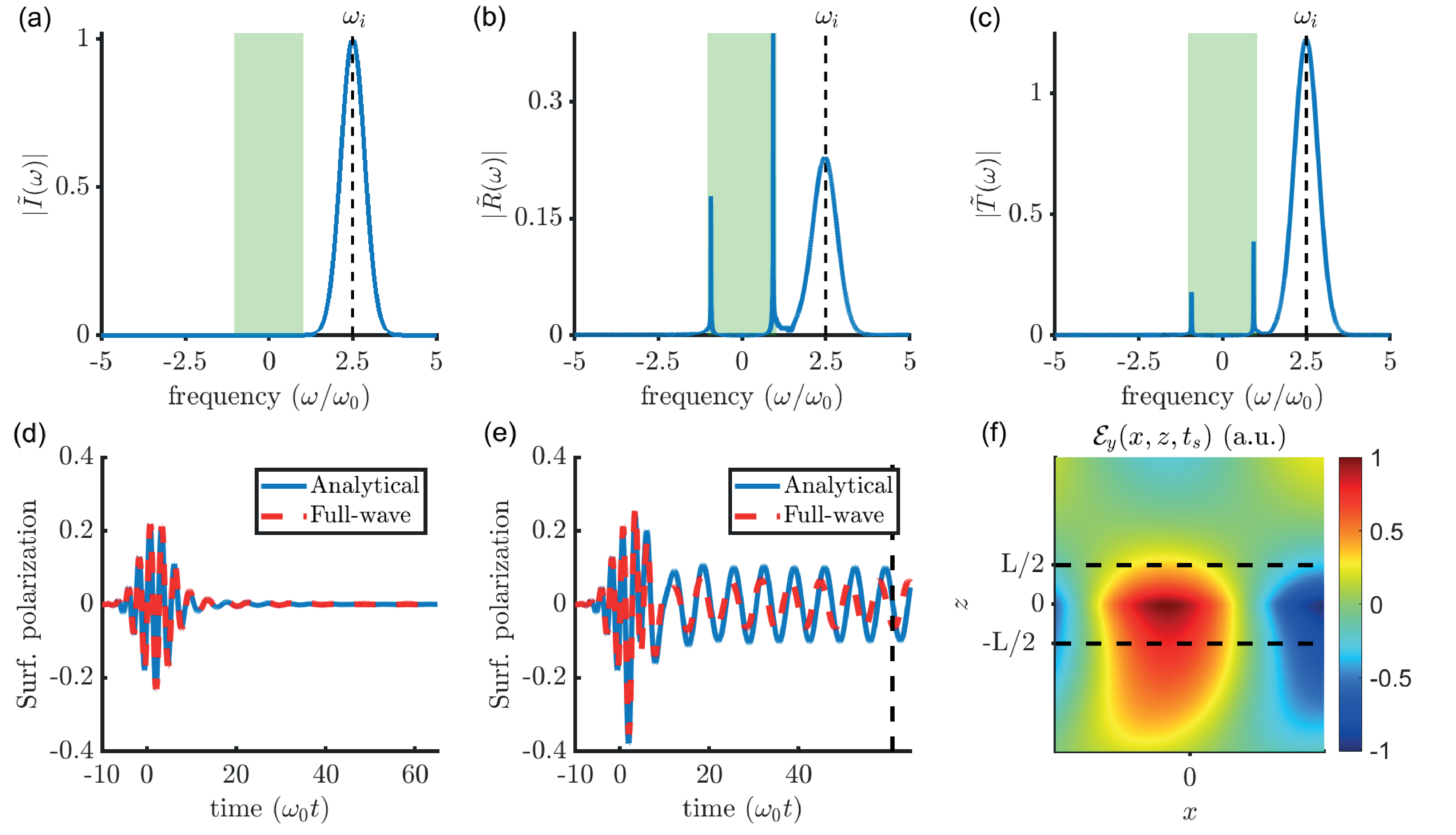}
	\caption{Example for oblique incidence ($k_x\simeq 1.44 \omega_0/c$), with $\varepsilon_L=2$, $\varepsilon_R=1$, $L=0.16 c/\omega_0$, $A_2=1.2$, $A_1=1$, and $\gamma=5 \cdot 10^{-3} \omega_0$.
 (a), (b), (c) Spectra of incident [$I(t)=E_i(0^-,t)/E_0$], reflected [$R(t)=E_r(0^-,t)/E_0$], and transmitted  [$T(t)=E_t(0^+,t)/E_0$] wave packets, respectively, with the dashed vertical lines indicating the position of the carrier frequency $\omega_i$, and the green-shaded region delimiting the evanescent range $|\omega|<\omega_L$.  All spectra are normalized with respect to the peak value of incident one.
(d), (e) Comparison between analytically computed [${\cal{P}}_y(x=0,t)/(\varepsilon_0 E_0 L)$]
and full-wave simulated [${\cal P}_y^{(s)}/(\varepsilon_0 E_0 L)$]  surface polarization
in the absence ($A_2=A_1=1$) and presence ($A_2=1.2, A_1=1$) of the temporal boundary, respectively. (f) Normalized electric-field map at time $t_s \simeq 61/\omega_0$, showing the excitation of a bound resonant mode (surface wave). The horizontal dashed lines delimit the thin layer interfaces at $z=\pm L/2$.}
	\label{fig4}
\end{figure*}

As another illustrative example, we consider a scenario where the wave packet is impinging at oblique incidence. Specifically, we assume $E_i(z=-L_1,t)=E_0 e^{-(t-t_0)^2/\sigma^2 - i\omega_{i} (t-t_0)}$ at a reference plane $z=-L_1 \simeq - 35 c/ \omega_0$, with $t_0 \simeq 53 /\omega_0$, $\sigma=4/\omega_0$, $\omega_i=2.5 \omega_0$, and $k_x \simeq 1.44 \omega_0/c$ (corresponding to an incidence angle of $24^{\circ}$ from the $z$-axis at the carrier angular frequency $\omega_{i}$). Moreover, we set $A_2=1.2$, $A_1=1$, $L=0.16 c/\omega_0$, $\varepsilon_L=2$, $\varepsilon_R=1$, and $\gamma=5 \cdot 10^{-3} \omega_0$.
Results are summarized in Fig. \ref{fig4}. In this case, the dispersion equation (\ref{eq:DE}) yields the angular eigenfrequencies $\Omega_{\pm} \simeq \pm 0.94 \omega_0$. As observed in the previous example, the incident field spectrum (Fig. \ref{fig4}a) is negligible at the resonance frequencies. Nevertheless, the reflection (Fig. \ref{fig4}b) and transmission (Fig. \ref{fig4}c) spectra exhibit sharp peaks (i.e., resonant modes) at these frequencies.
For the chosen parameters, these resonant modes are {\em bound} since their eigenfrequencies lie within the evanescent spectral range (highlighted with green shading in Figs. \ref{fig4}a-c). Therefore, the excitation of the natural resonance induced by the temporal boundary enables the coupling of a propagating wave packet with a bound mode (surface wave), without the need for any spatial subwavelength structuring. In contrast to the temporal grating introduced in \cite{Galiffi:2022po}, this effect is achieved without resorting to extended time-periodic modulations.

We validate the observed effect through full-wave simulations performed using two separate and distinct approaches (see \cite{supp} for details). Figures \ref{fig4}d,e compare the analytical surface polarization ${\cal P}_y(x=0,t)$ with the numerical counterpart. The agreement is very good, both in the absence and  presence of the temporal boundary. The slight discrepancy observed in Fig. \ref{fig4}e can be attributed to the finite-thickness effects, which are not accounted for in the analytical model. Importantly, in the presence of the temporal boundary, there is clear evidence of the excitation of a sinusoidal wave after the plasma frequency switch at $t>0$ (see Fig. \ref{fig4}e). 
For a clearer illustration, Fig. \ref{fig4}f shows an instantaneous distribution of the electric field after the plasma frequency switch, from which the excitation of a bound mode (surface wave) is apparent.

It is worth highlighting that the coupling with the bound resonant modes is governed by the amplitude coefficients $P_{\pm}$ in Eq. (\ref{P0}). This interaction is influenced not only by the temporal boundary but also by the overlap between the  incident wave packet spectrum and the Lorentzian resonant lineshape of the metasurface. Further details on this coupling, including the impact of detuning, are provided in \cite{supp}, along with the effects of imperfect synchronization between the impinging wave packet and the temporal boundary.

In conclusion, our study demonstrates that a temporal boundary in a dispersive metasurface can trigger the natural resonances of the system, leading to the coupling to radiative or bound modes. These modes, even in their bound form (i.e., surface waves), can efficiently couple with a propagating incident wave packet, without the need for spatial structuring or periodic temporal modulations. We have developed a semi-analytical framework, validated against rigorous full-wave numerical simulations, which provides useful insights into this phenomenon and enables its control. These outcomes may find broad applications in the field of compact and ultra-fast photonic devices. Importantly, the parameters chosen in our examples are in line with realistic values for semiconductor technological platforms operating at THz frequencies \cite{Lee:2018,Kamaraju:2014,Shi:2008}. For instance, in GaAs or Si slabs illuminated by a femtosecond optical pulse, a sizable plasma frequency change can occur within hundreds of femtoseconds (e.g., $t_A \simeq \sigma/10$ for a THz wave packet width $\sigma = 1$ ps).  Moreover, precise synchronization between an incident wave packet and a temporal boundary induced in a metasurface has been demonstrated in \cite{Lee:2018} by means of an optical-pump-THz-probe setup. Therefore, the experimental validation may be within reach, and will be further explored in forthcoming studies. 
Finally, we emphasize that our findings pave the way for future studies in quantum scenarios \cite{Ganfornina-Andrades:23} as well as in other wave systems. For example, our results could offer further physical insights into quantum quench dynamics, where a sudden change in a Hamiltonian parameter induces complex dynamics (see, e.g., \cite{quench1}). Moreover, our approach could be applied in acoustics by designing suitable space-time-varying metamaterials \cite{acou1,Wen2022}.
   
\section*{Acknowledgement}
C.R. acknowledges the funding support from the European Union -- NextGenerationEU under the Italian MUR National Innovation Ecosystem grant ECS00000041 - VITALITY - CUP E13C22001060006. 
M.A.V. thanks NATO Science for Peace and Security program (G5984) for partial financial support.
G. C. and V. G. acknowledge partial support from the University of Sannio via the FRA 2023 program.



%

\end{document}